\documentstyle{mn}

\input{epsf.sty}
\def\plotone#1{\centering \leavevmode
\epsfxsize=\columnwidth \epsfbox{#1}}

\def\msun{{\rm M}_\odot}
\def\rsun{{\rm R}_\odot}

\def\teff{T_{\rm eff}}
\def\dmc{{\dot M_{\rm crit}}}
\def\dd{{\rm d}}
\def\zr{\zeta_{\rm R}}
\def\RL{R_{\rm L}}
\def\m2i{M_{2 {\rm i}}}
\def\pcr{P_{\rm crit}}

\begin{document}

\title{Soft X--Ray Transients in the Hertzsprung Gap}
\author[Ulrich Kolb]{Ulrich Kolb$^{1,2}$\\ \\ 
$^1$ Astronomy Group, University of Leicester,
Leicester, LE1~7RH (uck@star.le.ac.uk) \\
$^2$ Max--Planck--Institut f\"ur Astrophysik, 
Karl--Schwarzschild--Str. 1, D--85740 Garching, Germany}

\maketitle
\begin{abstract}
We apply the disc instability model for soft X--ray transients
to identify system parameters along evolutionary sequences of black
hole X--ray binaries (BHXBs) that are consistent with transient 
behaviour. In particular, we focus on the hitherto neglected group of
BHXBs with intermediate--mass giant donor stars. These spend a
significant fraction of their X--ray active phase crossing the
Hertzsprung gap.    

Three Case B binary sequences with a black hole
accretor and $2.5\msun$ initial donor mass are presented in detail. 
We formulate rules which summarize the behaviour of these sequences
and provide an approximate description for Case B mass transfer in
intermediate--mass BHXBs. Chiefly, the timescale of the overall radius
expansion is given by the initial donor mass, while the surface
appearance is determined by the current donor mass.  

With these rules we obtain a general overview of transient and
persistent behaviour of all intermediate--mass BHXBs by just
considering single star sequences of different mass. 
We find that although systems in the process of crossing the 
Hertzsprung gap are in general persistently bright, with Eddington- or
super--Eddington transfer rates, there is a narrow instability strip
where transient behaviour is possible. This strip extends over a
secondary mass range $2.0 \la M_2 \la 3.5 \msun$. GRO~J1655--40 might
be such a system.  We predict that there are no BHXB
transients with donors more massive than $3.5\msun$, and no  
neutron star transients in the Hertzsprung gap.
\end{abstract}


\section{Introduction}

Low--mass X--ray binaries (LMXBs) are semi--detached compact binaries
with a black hole (BH) or neutron star (NS) primary that accretes mass
from an accretion disc. A Roche lobe filling low-mass main--sequence
or (sub)giant star feeds the disc. 
LMXBs appear in two main varieties: They are either
persistently bright, or have outbursts and long phases of
quiescence (e.g.\ White \& van~Paradijs 1996). The latter, known as
soft X-ray transients (SXTs; cf.\ Tanaka \& Shibazaki 1996 and Chen et
al.\ 1997 for reviews),
are believed to have an unstable accretion disc which alternates
between a hot and cool state (Cannizzo et al.\ 1982, Taam \& Lin 1984,
Cannizzo, Chen \& Livio 1995, King \& Ritter 1998), similar to dwarf
novae among cataclysmic variables.  

A simple criterion for stable disc accretion is 
that the temperature $T_d$ at the outer disc edge is larger than the
hydrogen ionization temperature $T_H \simeq 6500~K$ (e.g.\ King,
Kolb \& Burderi 1996; hereafter KKB). 
As $T_d$ increases with the mass transfer rate $\dot M$, a system is
persistently bright if $\dot M$ is larger than the critical rate 
$\dmc$ where $T_d=T_H$. The system is transient if $\dot M < \dmc$. 

An important difference between dwarf novae and SXTs is the dominant
role irradiation of the accretion disc by the central accreting source
plays in LMXBs. Van Paradijs (1996) showed that the observed critical
transfer rate separating transient from persistent neutron star LMXBs
is much lower than the critical rate separating dwarf novae from
persistently bright (novalike) cataclysmic variables. This can be
understood quantitatively by assuming that irradiation dominates over
viscous heating at the outer disc rim, hence stabilizing the disc even
at a smaller mass transfer rate. Similarly, irradiation naturally
explains the observed large ratio of optical to X--ray flux from the
disc and the observed long, quasi--exponential decline after an SXT outburst
(King \& Ritter 1998).   

Models for the evolution of LMXBs predict the mass transfer rate 
and thus by comparison with $\dmc$ the appearance of the system as a
transient or persistently bright X--ray source
(assuming that the instantaneous transfer rate is close to the
evolutionary mean transfer rate).
This is a powerful
diagnostic tool for testing evolutionary models, or, in turn, the disc
instability model for SXTs. King, Kolb \& Szuszkiewicz (1997;
hereafter KKS), King et al.\ (1997) and KKB investigated short--period
systems with essentially unevolved donors, and long--period systems
having low--mass giant donors with degenerate helium cores. 
In the following we extend this work, and present evolutionary models
for black hole X--ray binaries (BHXBs) with intermediate--mass donors
in the process of crossing the Hertzsprung gap, a hitherto largely
neglected parameter range. 
We explore to what extent such systems appear as SXTs.
A major motivation for this study is provided by recent optical
observations of the transient X--ray source GRO~J1655--40 (Orosz \& Bailyn
1997), well--known from the apparent superluminal motion in its
radio jets (Hjellming \& Rupen 1995). System parameters derived for
GRO~J1655--40 place this BH binary just in the above parameter
range dicussed here. Some of the results presented below have already
been applied to a discussion of the evolutionary state of
GRO~J1655--40 (Kolb et al.\ 1997).  

We put intermediate--mass BHXBs into context by reviewing main
concepts of LMXB evolution briefly in Sect.~2. In Sect~3 we present
in detail three evolutionary sequences, 
with parameters close to the ones observed in GRO~J1655--40. 
We generalize the results of these calculations in Sect.~4 in the form
of three
rules which approximately describe the general evolution of
intermediate--mass BHBXs. These rules allow one to obtain a general
overview of transient and persistent behaviour of all
intermediate--mass BHXBs by just considering single star 
sequences of different mass. 
Sect.~5 concludes with a critical discussion.

\section{Evolution of LMXBs}

The mechanism driving mass transfer naturally divides LMXBs into two
distinct classes.
In the first group, nuclear expansion of the
secondary maintains the semi--detached state, while in the second group,
orbital angular momentum losses such as gravitational radiation and
magnetic braking drive mass transfer. For convenience we denote
neutron star or black hole LMXBs in the two groups with n--driven or
j--driven systems.   

Generally, n--driven LMXBs are long--period systems with (sub)giant
donor stars on the first giant branch. Hydrogen burns in a shell
source above a nuclearly inactive helium core. The orbital period
increases until mass transfer terminates when core helium
burning ignites, or when the hydrogen--rich envelope is fully
transferred to the primary, whichever happens earlier. The final state
is a wide, detached binary with a white dwarf secondary and a BH or
NS primary (possibly a millisecond pulsar). The LMXB lifetime depends
on the initial secondary mass and orbital distance and can be as long
as $10^9$~yr if both are small, but is typically $10^7 - 10^8$~yr. 

Conversely, j--driven systems are short--period binaries with
main--sequence donors and evolve to shorter orbital period.   
If the j--driven evolution is the same as for cataclysmic variables
(see e.g.\ Kolb 1996, King 1988 for reviews) the bright phase where 
magnetic braking operates ($P \ga 3$~h) typically lasts
$10^8$~yr. For $P \la 3$~h mass transfer continues indefinitely 
at a lower rate,
but the period increases again when the secondary becomes
degenerate, i.e.\ when its mass is $\la 0.06\msun$.

A critical bifurcation period $P_b \simeq 1-2$~days (Pylyser \&
Savonije  
1988, 1989) separates j--driven and n--driven LMXBs. Systems born with
$P<P_b$ evolve towards shorter periods, and those born with $P>P_b$
towards longer $P$.   
Similarly, the initial donor mass $\m2i$ plays a role in determining
the group membership:
If $\m2i \la 0.8\msun$ the system is j--driven, as the donor's
main--sequence lifetime is longer than a Hubble time. If $\m2i \ga
1.5 \msun$ the system cannot evolve to shorter orbital periods as 
magnetic braking does not operate in such massive stars, and
gravitational radiation alone is too weak to dominate the evolution. 
In the intermediate range $0.8 \la \m2i \la 1.5$ both groups exist. 
Mass transfer stability defines an upper limit for the mass ratio
$q=M_2/M_1$ ($M_2$ is the donor mass, $M_1$ the primary mass), hence
for $M_2$ and $\m2i$. 
This limit depends on the donor's response to mass loss
(Hjellming 1989), i.e.\ is a non--trivial function of its 
evolutionary state. As a rule, the limit is roughly $M_2 \la M_1$ 
in the case of conservative mass transfer, but can be somewhat larger
with mass loss from the system (cf.\ Kalogera \& Webbink 1996, where
the NS case is discussed in detail). If the donor is more massive 
the system suffers a short, violent phase of mass transfer/loss and is
unlikely to appear as an X--ray source.   
Obviously, 
neutron star X-ray binaries with Roche--lobe filling donors more
massive than $2-3\msun$ do not exist, 
while there is no such upper
limit for BHXBs, 
hence no clear separation between low--mass and high--mass BHXBs.
The importance of BHXBs with intermediate--mass donors
was first pointed out by Romani (1994).

Recently, two limiting cases of LMXB evolution have been considered  
to determine the incidence of transient behaviour among LMXBs: 
KKB and KKS described the j--driven evolution of completely unevolved
(ZAMS) donors. They found that BH systems are always transient,
consistent with observations, and that donor stars in NS systems need
to be very close to the end of core H burning in order to be
transient, consistent with the observed small fraction of transient NS
LMXBs. King et al.\ (1997) and KKS studied n--driven LMXBs with 
low--mass donor stars  well--established on the first giant branch,
and found that essentially all of them are transient.
They considered giant donors with a thin H burning shell source
above a degenerate He core where the stellar radius $R$ and luminosity
$L$ are unique functions, $R \propto M_c^{5.1}$, $L \propto
M_c^{8.1}$, of the slowly growing core mass $M_c$, essentially
independent of the total stellar mass (e.g.\ Webbink, Rappaport \&
Savonije 1983). Stars with mass $\la 2\msun$ indeed establish such a
structure soon after the end of central H burning and spend a
considerable time ascending the first giant branch, before the
ignition of helium burning in the centre terminates
their radius expansion. Most n--driven NS LMXBs are well described in
this way, but not BHXBs with more massive donor stars ($M \ga 2.0 -
2.5\msun$).  These ignite core helium burning before
the core becomes highly degenerate, i.e.\ before the first giant
branch evolution governed by the above core mass relations takes
hold. Simplified descriptions for the evolution of these stars fail,
and full stellar models are needed. 

We consider these n--driven intermediate--mass BHXBs in the next
sections.

\section{Illustrative Case B Mass Transfer Sequences With Black Hole
Accretors}    

The binary evolutionary phase we consider here is Case B mass
transfer: the donor star fills its Roche lobe  after termination of
core H burning but before the ignition of core He burning. 
Case B mass transfer is well--studied (e.g.\ Kippenhahn et
al.\ 1967, Pylyser \& Savonije 1988, Kolb \& Ritter 1990, De~Greve
1993) in the context of Algol evolution. All these studies assume that
the donor is more massive than the accretor, simply because the faster
evolving more massive component always fills its Roche lobe first. 
In BHXBs, with a more complicated evolutionary history,
Case B mass transfer begins with an already inverted mass
ratio, i.e.\ the donor is less massive than the BH. Hence there is no 
initial rapid phase with extremely high transfer rate.

In the following we present calculations of three Case B mass
transfer sequences on to a BH primary from a secondary with initial
mass close to the transition region between degenerate and
non--degenerate helium ignition. We report briefly on the
numerical technique and associated problems (Sect.~3.1), describe
in detail the evolution of the secondary star (Sect.~3.2), and then
focus on the resulting appearance in X--rays along the sequences in
Sect.~3.3.

\subsection{Input Parameters and Computational Technique} 

We used Mazzitelli's stellar evolution code in a version as described
by Mazzitelli (1989; see also references therein) with pre--OPAL
opacities. The computations started from chemically homogenous ZAMS
models with Population I mixture ($X=0.70$, $Y=0.28$). Convection is
treated by the standard mixing length theory; a calibration to a solar
model determines the mixing length parameter to $1.4$ (no overshooting
was allowed).  

To allow the application to binary evolution 
the mass transfer rate $\dot M$ was calculated for each
time step according to 
\begin{equation}
  \dot M \propto  \exp \left( \frac{\Delta R}{k H} \right)
\label{eq1}  
\end{equation}
(cf.\ Kolb \& Ritter 1990), 
where $\Delta R = R - \RL$ is the difference between the secondary's
radius $R$ and Roche lobe radius $\RL$, $H$ the atmospheric pressure scale
height, and $k$ a numerical constant, usually taken as $k=1$. Although
(\ref{eq1}) is strictly valid only for $\Delta R < 0$ and has to be
replaced by an expression with an approximate power--law dependence on
$\Delta R/H$ for $\Delta R > 0$, we use (\ref{eq1}) for any $\Delta
R$. Furthermore, to avoid numerical instabilities at high
transfer rates we set $k=10$ for $\dot M \ga 10^{-7} \msun$
yr$^{-1}$.
This procedure is justified as long as $\Delta R/R \ll 1$ (Kolb \&
Ritter 1990), i.e.\ as long as the stellar radius is close to the
critical Roche radius. For the sequences presented here this is
always the case. A high--resolution reference sequence with $k=1$ was
calculated to check the validity of sequences obtained with $k=10$
explicitly in the case of sequence S3 (see below). 
Attempts to retain the damping factor $k=10$ also in later phases
(where $\Delta R<0$) failed as $H/R$ becomes non--negligible for
extended giants. A damped evolution would proceed qualitatively
differently from the true evolution with $k=1$, leading to higher mass
transfer rates and a correpondingly earlier termination of mass
transfer.

\subsection{Detailed Description of the Sequences}

\begin{figure}
\plotone{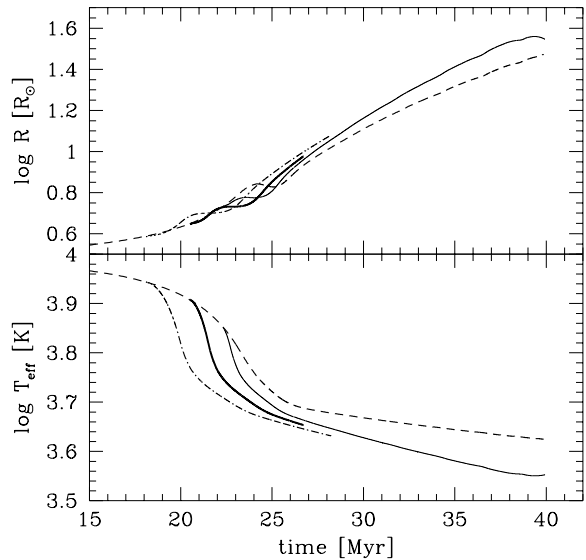}
\caption{
Radius $R$ (upper panel) and effective temperature $\teff$ (lower
panel) of a $2.5\msun$ single star (dashed) and
the secondary star in the binary sequences S1 (dash--dotted), S2
(thick solid line) and S3 (thin solid line), 
cf.\ Tab.~1, as a function of time elapsed since the (donor) star left
the main sequence. 
}
\end{figure}

\begin{figure}
\plotone{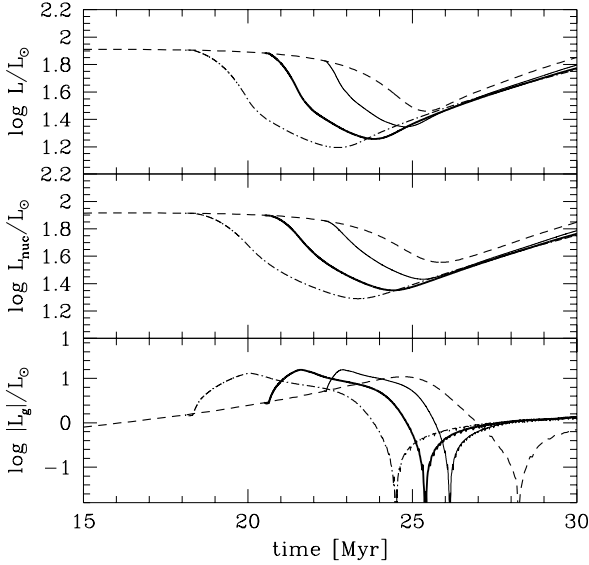}
\caption{
Surface luminosity $L$ (top), nuclear luminosity $L_{\rm nuc}$ (middle
panel) and absolute value of the gravo-thermal luminosity
$L_g=L-L_{\rm nuc}$ (bottom) of a $2.5\msun$ single star 
(dashed) and the secondary star in binary sequences
S1--S3, as a function of time elapsed since the
(donor) star left the main sequence.  The linestyle is as in Fig.~1.
}
\end{figure}

\begin{table}
\caption{Binary sequence parameters: BH mass $M_1$, donor mass
$M_2$, orbital separation $a$; the last columns lists an internal
sequence number.}
\begin{tabular} {lllll}
sequence & 
$M_1/\msun$ & $M_2/\msun$ & $a/\rsun$ & int \\
\hline
S1 (initial) & 6.8 & 2.5 & 13.2 & b860 \\
S2 (initial) & 6.8 & 2.5 & 15.2 & b849 \\
S3 (initial) & 6.8 & 2.5 & 18.5 & b830/32 \\
\hline
S3 (final)   & 8.726 & 0.574 & 213.1 & \\
\hline
\end{tabular}
\end{table}

\begin{figure}
\plotone{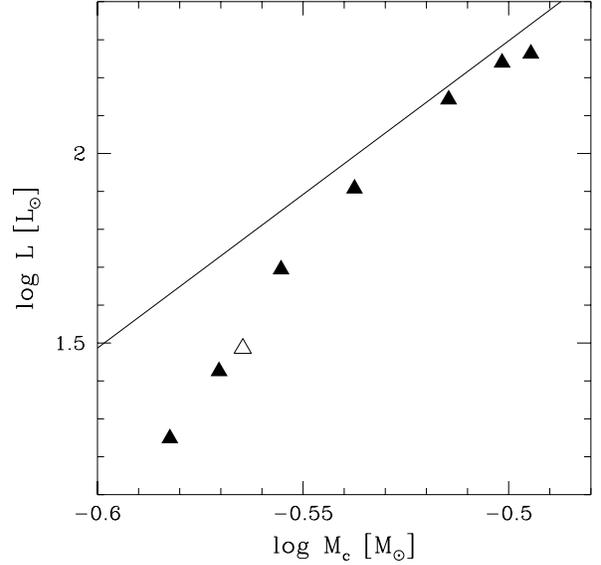}
\caption{
Surface luminosity $L$ as a function of core mass $M_{\rm c}$ (defined
as the mass enclosed by the shell where the H burning energy generation
rate is maximal). Full triangles: selected models along sequence
S3. The triangles with smallest and largest $M_{\rm c}$ correspond to 
models at the luminosity minimum and close to central He ignition,
respectively. The open triangle marks a $2.5\msun$ single
star at its luminosity minimum. The positions of models with larger
$M_{\rm c}$ along the single star sequence coincide with the positions of
sequence S3. The solid curve is the core--mass luminosity
relation for low--mass 
($< 1.5 \msun$) 
giant stars, cf.\ King et al.\ 1997.
}
\end{figure}

All three binary sequences, S1--S3, begin mass transfer with a
$2.5\msun$ donor star and a $6.8 \msun$ black hole. The initial
orbital separation is $a_i = 13.2$ (S1), $15.2$ (S2) and $18.5 \rsun$
(S3), see the summary in Tab.~1. 
Mass transfer is conservative, i.e.\ the total binary mass and
orbital angular momentum is constant. Sequence S3 follows the donor
star until ignition of central helium burning. The subsequent
evolution is detached, and we evolved the secondary further through
central helium burning and the brief subsequent asymptotic giant
branch phase 
with weak thermal pulses. In contrast, sequences S1 and S2 have been
terminated once the secondary was established on the first giant
branch.

In Figs.~1 -- 5 we show characteristic parameters along the sequences
S1--S3, sometimes together with the evolution of a $2.5 \msun$ single
star (for convenience referred to as sequence S0). 

Figure~1 (upper panel) reveals that the overall radius evolution with
time is very similar in all sequences S0--S3, except for the ``plateau
phase'' soon 
after mass transfer turn--on, which corresponds to a brief radius
contraction phase at the red end of the Hertzsprung gap of S0. 
This phase occurs at a characteristic radius, the ``plateau radius''
$R_p$, and  
will be examined in more detail below. The similarity
shows that significant mass loss from the outer envelope has little
effect on the overall expansion of the donor star, which is governed
by the nuclear evolution of the core region. The overall
temporal evolution of the surface luminosity $L$ is systematically
different for sequences with and without mass transfer (Fig.~2): in
S1--S3 mass 
loss causes a steep build-up of gravo--thermal luminosity $L_g$ ($L_g
< 0$) and a slight reduction of 
the nuclear luminosity $L_{\rm nuc}$ from the hydrogen shell source.
This behaviour is superimposed on the luminosity evolution pattern of S0 
in the Hertzsprung gap, where hydrogen burning changes from a thick to
a thin shell. The net effect is that thin hydrogen shell burning is
established earlier, at a slightly smaller core mass. Once 
thin H burning is fully established the luminosity grows with further
growing core mass. The corresponding effective core mass--luminosity
relation is much steeper than the standard relation $L \propto M_c^{8.1}$
for fully degenerate cores, and meets the latter one only immediately
before core helium ignition (Fig.~3). 
As a result, the secondary in S1--S3 begins its ascent along the
first giant branch at a significantly smaller surface luminosity than
the $2.5 \msun$ single star. The drop of $L$, a familiar feature of
Case B mass transfer, goes along with a
correspondingly fast decrease of $\teff$ to $\log \teff/{\rm K} \simeq
3.70$, i.e.\ the secondary in the mass loss sequence crosses the
Hertzsprung gap even faster than the $2.5 \msun$ single star
(Fig.~1, lower panel). 
Roughly, at any time along S1--S3 in the Hertzsprung gap, $L$ and
$\teff$ are the same as for a single star with the same mass and
radius as the secondary star at that time.

With the ignition of helium burning the secondary's radius decreases
sharply and mass transfer terminates. At this point the binary
parameters of S3 are $M_1 = 8.726 \msun$, $M_2 = 0.574 \msun$
(with core mass $0.320 \msun$) and $P = 118.3$~days, $a = 213.1\rsun$
(see Tab.~1). For the next $270$~Myr the secondary burns helium in the
centre 
and is well inside its Roche lobe ($R\simeq 8-10\rsun$ versus
$\RL=34.6\rsun$). Subsequently, He burning moves to a shell source and
the radius increases again, at a core mass of $\simeq 0.420
\msun$. The ensuing double shell burning is unstable and leads to
thermal pulses. We find that these pulses are rather weak, with
decreasing maximum radius. The radius maximum in subsequent pulses is
always smaller than the Roche radius (though not very much), and quickly
decreasing. Eventually the secondary contracts towards the white dwarf
stage. If there are no significant wind losses during the thermal
pulses the final state of S3 is a wide, detached binary with a $\simeq
9\msun$ black hole primary, a $\simeq 0.55\msun$ carbon--oxygen white
dwarf companion and $\simeq 5.6$ months orbital period.

\subsection{Transient Versus Persistent Accretion}

\begin{figure}
\plotone{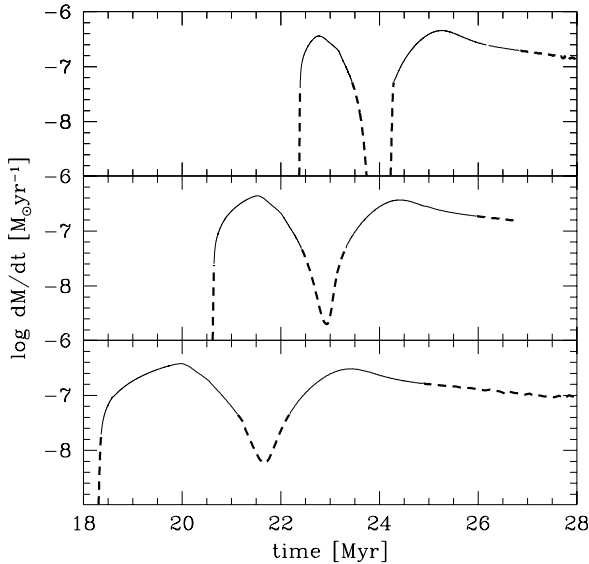}
\caption{
Mass transfer rate $\dot M$ as a function of time (elapsed since the
star left the main sequence) for sequences S1 (bottom panel), S2
(middle panel) and S3 (upper panel). The linestyle is broken when the
mass transfer rate is smaller than the critical rate $\dmc$ for disc
instability (taken from KKS).
}
\end{figure}

\begin{figure}
\plotone{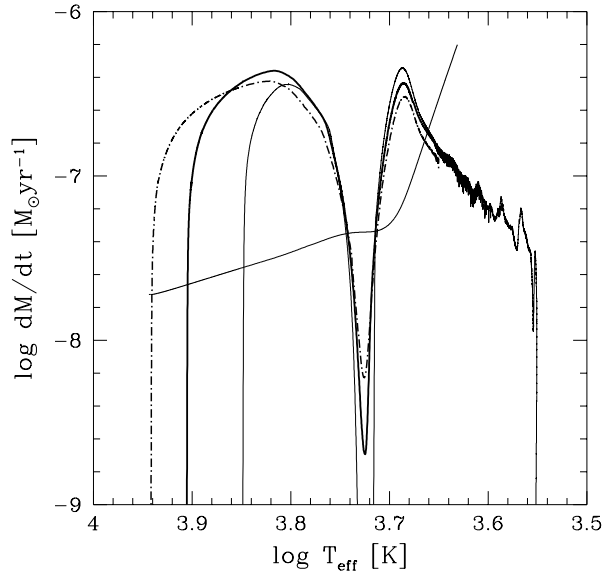}
\caption{
Mass transfer rate $\dot M$ as a function of effective temperature
$\teff$ for sequences S1 (dash--dotted), S2 (thick solid line) and S3
(thin solid line). Also shown is the critical rate for disc
instabilities (from KKS) for sequence S1. 
}
\end{figure}

We distinguish transient from persistent BHXBs by comparing the mass
transfer rate $\dot M$ with the critical rate $\dmc$ for
which hydrogen just ionizes at the outer disc edge. A system is
transient if $\dot M < \dmc$. In Fig.~4 we plot $\dot M$
as a function of time along sequences S1--S3. The linestyle is broken
if $\dot M < \dmc$, with $\dmc$ taken from KKS.
Obviously, the accretion disc is stable for about $\simeq 4-5$~Myr
when the secondary crosses the Hertzsprung gap, and the   
transfer rate is close to (or slightly above) the Eddington limit
\begin{equation}
\dot M_{\rm edd} \simeq 2 \times 10^{-8} \msun {\rm yr}^{-1} \:
\left( \frac{M_1}{\msun} \right) 
\label{eq2}
\end{equation}
(assuming an effective accretion efficiency $\eta=L_{\rm acc}/\dot M
c^2=0.1$). 
The system becomes a transient source soon after it
begins to ascend along the Hayashi line, and remains transient until
helium ignites in the centre $\simeq 12$~Myr later. Significantly,
there is also a short transient phase (lasting $\simeq 1$~Myr) right
{\it in} the Hertzsprung gap, centered around $\log \teff/{\rm K} =
3.73$. This phase is caused by the transition from thick to thin shell
source burning, the same effect that leads to a brief
radius contraction when the $2.5 \msun$ single star crosses the
Hertzsprung gap. The associated change of the internal structure of
the star is highly non--linear and complex. Just as the ultimate
cause for the expansion to the red giant state cannot be easily
derived from a simple analysis of the stellar structure equations,
there is no easy way to understand this temporary halt in the
expansion either. 

The Hertzsprung gap transient phase in S1--S3 and the
radius contraction phase of S0 all occur at the same range of
$\teff$ (Fig.~5), while the corresponding plateau radius $R_p$ is
largest for 
S0, and progressively smaller the earlier mass transfer began in the
Hertzsprung gap. The secondary mass in the plateau
phase is $1.9$, $2.05$ and $2.25 \msun$ for sequences S1, S2 and S3,
respectively. In fact, $R_p(M_2)$ is practically the same as the 
plateau radius of a single star with the secondary's mass in the
plateau phase.

Details of the exact location (e.g.\ the radius $R_p$), duration and depth
of the radius contraction phase of a single star must
necessarily depend on details of the stellar input physics, in
particular on opacities and treatment of convection. 
As an example, the $2.5\msun$ single star track obtained by Salaris et
al.\ (1997) with OPAL opacities shows a much less pronounced radius
contraction phase at a slightly higher effective temperature ($\log
\teff/{\rm K} \simeq 3.75$) than our sequence S0. 
Hence we do not expect our models to reproduce the precise location of
the real transient phase in the Hertzsprung gap, and the actual mass
transfer rate in this phase. But our sequences certainly show the
differential 
change of that phase with varying initial orbital distance: the
earlier mass transfer starts in the Hertzsprung gap, the less
pronounced the 
decrease of $\dot M$ in the transient phase. If the secondary is 
already very close to the plateau phase when it fills its 
Roche lobe for the first time, the system might almost detach in the
transient phase and not appear as an X-ray source at all; but if the system
begins mass transfer early in the Hertsprung gap it certainly will.

\section{Generalization: Intermediate--Mass BHXB Evolution}

We formulate three rules which summarize the behaviour of the above
evolutionary sequences and provide an approximate description for the
Case B evolution of intermediate--mass BHXBs. 

\begin{quote}
(R1) the overall time evolution $R(t)$ of the donor star radius in a
binary sequence with initial donor mass $\m2i$ is the same as
the time evolution of the radius of a single star with mass
$M_0=\m2i$ 
\end{quote}

\noindent while

\begin{quote}
(R2) the plateau phase (the brief contraction phase or slow--down of
the expansion phase)
in the Hertzsprung gap occurs at a donor radius $R_p$ equal to the
plateau radius of a single star with the donor's mass
\end{quote}

\noindent
\begin{quote}
(R3) a donor star with mass $M_2$ and radius $R$ has the same 
effective temperature $T_{\rm eff}$ (and surface luminosity $L$) as
a single star with mass $M_0=M_2$ and radius $R$
\end{quote}

In other words: the timescale of the overall radius expansion is given
by the initial donor mass, while the surface appearance and the
plateau phase is given by the current donor mass.

In the following we use these rules to investigate the incidence of
transient and persistent behaviour in intermediate--mass BHXBs with
arbitrary initial mass and initial separation, simply by considering
the evolution of single stars in this mass range. 

In order to do so we note that in the case of stationary mass transfer
($\dot R/R 
= \dot \RL/\RL$) the transfer rate is given by  
\begin{equation}
 \dot M = M_2 \: \frac{K}{\zeta_{\rm eq} -\zr} \,
\label{eq3}
\end{equation}
(cf.\ e.g.\ Ritter 1996).
Here $K = (\dd \ln R/\dd t)_{M={\rm const.}}$ is the radius expansion
in the absence of mass loss (e.g.\ due to nuclear evolution),
$\zeta_{\rm eq}$ the thermal equilibrium mass--radius exponent, and
$\zr$  the  Roche 
lobe index ($\zr \simeq 2M_2/M_1 - 5/3$ for conservative
evolution). We set $\zeta_{\rm eq} = 0$ and, using (R1),  
approximate $K$ by $\dd \ln R/\dd t$ along the single star sequence.
It is clear that these secondaries are not in thermal equilibrium
(Fig.~2), and their radius expansion is not purely nuclear. However,
the mass loss  perturbs the overall radius evolution only slightly. 

\begin{figure}
\plotone{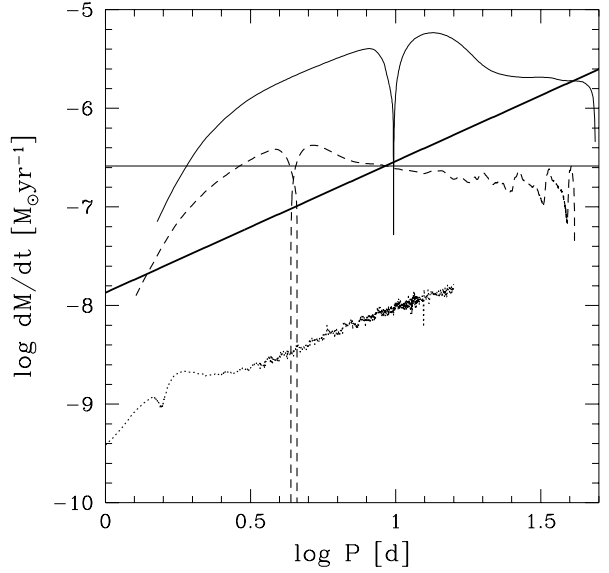}
\caption{
Transfer rate $\dot M_{\rm i}$ versus orbital period $P_{\rm i}$ at
turn-on of mass transfer, for three different donor masses (full line:
$3.5\msun$; dashed: $2.5\msun$; dotted: $1.3\msun$), estimated from
Eq.~(\protect{\ref{eq3}}) and assuming a $8\msun$ primary. The
horizontal line indicates the corresponding Eddington transfer rate,
see (\protect{\ref{eq2}}). The thick solid line is the critical mass
accretion rate $\dmc$ separating transient ($\dot M < \dot
M_{\rm cr}$) from persistent ($\dot M > \dot M_{\rm cr}$) systems,
taken from KKS. 
}
\end{figure}

Then we obtain the approximate binary
evolution for given initial donor mass $\m2i$ by integrating $\dot
M(t)$, i.e.\ $K(t)$ of a {\em single star} with mass $M_0=\m2i$, over
time. To decide the stability of the accretion disc we need to
know the mass transfer rate $\dot M$ as a function of orbital period $P$. 
The main characteristics of the real evolutionary tracks $\dot M (P)$
are already evident from Fig.~6 where we plot the mass transfer rate $\dot
M_{\rm i}$ {\em soon after turn--on of mass transfer}, estimated from
(\ref{eq3}), as a function of {\em initial} orbital period $P_{\rm
i}$, for secondaries in the mass range of interest ($\m2i = 1.3 - 3.5
\msun$). Roche geometry determines the initial period to $\log (P_{\rm
i}/{\rm d}) = 1.5 \log (R/\rsun) - 0.5 \log (\m2i/\msun) - 0.433$. 
Also shown is the critical rate $\dmc$ for disc instability, taken
from KKS, and the Eddington accretion rate (\ref{eq2}) for a typical
$8\msun$ BH accretor. 

For $\m2i \la 1.9 \msun$ we find that $\dot M_{\rm
i} < \dmc$ at any $P_{\rm i}$, while for $\m2i \ga 1.9 \msun$ the 
transfer rate is smaller than $\dmc$ either if the donor is in the
plateau phase, $P_{\rm i} = P_{\rm i}(R_p(\m2i),\m2i)=P_p(\m2i)$, or if 
$P_{\rm i}$ is sufficiently long, i.e.\ longer than a
critical period $\pcr$ where $\dot M_{\rm i} = \min (\dmc,\dot 
M_{\rm edd})$. The Eddington rate enters because for $\dmc > \dot
M_{\rm edd}$ the mass {\em
accretion} rate in any system with super--Eddington mass transfer rate 
is probably limited to a value near $\dot M_{\rm edd}$, 
hence the accretion disc is unstable even if $\dot M > \dmc$, 
(assuming that the character of disc irradiation does not change for
super--Eddington mass transfer;  
note that because $\dmc$ is determined by irradiation, it is the
central accretion rate which enters the stability criterion.)

\begin{figure}
\plotone{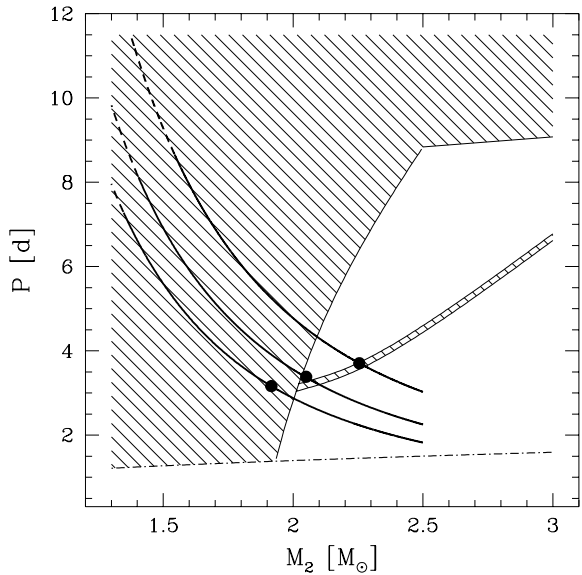}
\caption{
Orbital period -- secondary mass ($P-M_2$) plane for BHXBs, 
showing the exclusion zone for transients (unhatched). A BH
mass of $8\msun$ was assumed. 
Systems in the unhatched region are always persistent; systems in the  
narrow hatched instability strip are always transient. Systems which
are born in the large hatched area are always transient, but systems
which have evolved from the unhatched into the large hatched area
will remain persistent for some time before they, too, 
become transient. Evolutionary tracks for sequences S1--S3 are also
shown (solid where persistent, dashed where transient). The heavy dot
along each sequence marks the location of the transient phase in the
Hertzsprung gap.
The dash--dotted line is the binary period when the secondary
fills its Roche lobe on the turn--off main--sequence. 
}
\end{figure}

\begin{figure}
\plotone{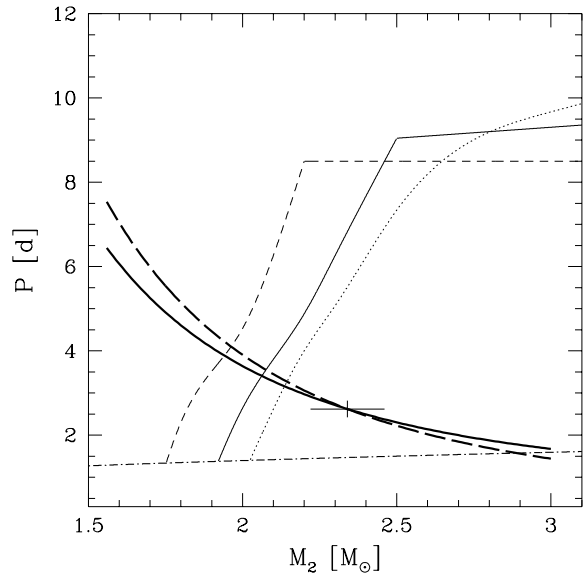}
\caption{
As Figure 7, but with the critical boundary between transient and
persistent systems for different BH masses
(dashed: $5\msun$; solid: $10\msun$; dotted: $15\msun$).  
The cross indicates the observed
system parameters of GRO~J1655--40 (Orosz \& Bailyn 1997). The solid
curves represent evolutionary tracks assuming conservative evolution
(full line) and evolution with constant black hole mass (mass leaving
the system carries the black hole's specific orbital angular momentum;
dashed).  
}
\end{figure}

We show $\pcr$ as a function of $\m2i$ in a plot of orbital period
versus donor mass (Fig.~7; a similar figure was already published in Kolb
et al.\ 1997). This critical period separates BHXBs with stable
accretion disc (unhatched region) from systems with unstable discs
(hatched region) at turn--on of mass transfer. The narrow hatched
``instability strip'' $P_p(\m2i)$ bisecting the unhatched region is the
plateau phase.
This strip terminates at $\m2i \simeq 3.5\msun$, as for more massive
donors $\dot M > \dmc$ even when the radius expansion slows.

We now consider the location of evolutionary sequences in Fig.~7, and
discuss the relevance of $P_p$ and $\pcr$ for them. After turn--on of
mass transfer a binary evolves to smaller $M_2$ and longer $P$. In the
case of conservative evolution the evolutionary tracks follow curves
with $(M-M_2)^3 M_2^3 P =$~const.\ ($M$ is the total binary mass), see
the tracks of sequences S1--S3. In binaries with larger initial
donor mass ($\m2i \ga 2.5 \msun$) the transfer rate is
super--Eddington and the evolution  
non--conservative. The primary accretes at $\dot M_{\rm edd}$, and
surplus material leaves the system at a rate
$\dot M - \dot M_{\rm edd}$, carrying some specific angular momentum
(of order the BH's specific orbital 
angular momentum) determined by details of how and where
the material is accelerated. In this case the mass transfer rate is
slightly lower and the period increase significantly faster than   
for conservative evolution, i.e.\ the evolutionary tracks 
would have a slightly larger slope $|\dd P/\dd M_2|$ 
(cf.\ the example in Fig.~8).  

A simple corollary of (R2) is that a system will reach the transient
plateau phase when its evolutionary track crosses the 
instability strip in Fig.~7. This is because $P_p$ is a unique
function of the donor mass, independent of the initial donor mass. 
In contrast, as the initial donor mass dictates the mass transfer rate
outside the plateau phase, we can expect that the critical period where
the system changes from a persistently bright to a transient X--ray
source is close to $\pcr(\m2i)$. Then, 
as a consequence of the opposite slopes of evolutionary tracks and the
curve $\pcr(\m2i)$ in Fig.~7, a system remains persistently bright
when its track leaves the unhatched region, and becomes transient only
later when its period roughly equals $\pcr(\m2i)$. 
The sequences S1--S3 (solid where persistent, broken
where transient; heavy dots mark the plateau phase)
and actual integrations of (\ref{eq3}) confirm this.
Hence no system in the unhatched region is transient, it
is a ``transient exclusion zone'', bisected by the Hertzsprung gap
instability strip. The hatched region, on the other hand, may contain
both persistently bright and transient systems. 

Fig.~8 shows how the transient exclusion zone depends on the BH 
mass $M_1$. As $M_1$ increases the steep branch of $\pcr(M_2)$ 
moves to larger $M_2$, while the flat branch moves to longer $P$. 
The BH mass does not affect the location of the instability strip.

\section{Summary and Discussion}

In extension of the work by KKB, KKS and King et al.\ 1997
we studied the evolution of BHXBs with intermediate--mass
giant donor stars in the context of the disc instability model
for soft X-ray transients. 
These represent a hitherto somewhat
neglected group of systems with very high mass transfer
rates and donors spending a significant fraction of the X--ray active
time in crossing the Hertzsprung gap. 
We find that although systems in the process of crossing the
Hertzsprung gap are 
in general persistently bright with Eddington- or super--Eddington 
transfer rates, there is a narrow instability strip where transient
behaviour is possible. This strip extends over a
secondary mass range $2.0 \la M_2/\msun \la 3.5$ and is roughly given by 
$P/{\rm d} = 4.3 M_2/\msun - 6.2$, but the precise location is subject
to uncertainties in the stellar input physics. 
We predict that BHXBs in the Hertzsprung gap are not transient if
the donor is more massive than $\simeq 3.5\msun$, and that 
neutron star LMXB transients in the Hertzsprung gap do not exist.
The latter is a consequence of a number of factors: in neutron star
LMXBs the upper (stability) limit for the donor mass is smaller, the
mass transfer rate is larger as the systems are closer to mass transfer
instability, and the critical rate $\dmc$ is possibly lower by almost
an order of magnitude (see KKS).   

In our considerations we relate the evolutionary state of a system
to the stability of its accretion disc, hence its appearance 
in X--rays, by comparing the evolutionary mean mass transfer
rate $\dot M$ to the critical transfer rate $\dmc$ for disc
instability. Both the identification of the actual, instantaneous mass
transfer rate with $\dot M$ and the normalisation of $\dmc$ are
subject to some uncertainties:

If the actual mass transfer rate deviated from the
calculated evolutionary mean the system could of course be transient
where it is predicted to be persistently bright, and vice
versa. Although such deviations 
from the secular mean rate are quite common in CVs (e.g.\ Warner 1987)
there is no observational evidence for a similar large scatter of the
mass transfer rate at a given orbital period in LMXBs (cf.\ 
van~Paradijs 1996; Chen, Shrader \& Livio 1997). Also, the most
promising theoretical explanation for the observed spread of $\dot M$
in CVs is that the systems undergo mass transfer limit cycles caused
by weak irradiation of the secondary (Ritter, Zhang \&
Kolb 1995; King et al.\ 1996). In LMXBs this mechanism does not work
as the irradiating flux on the secondary is too large (cf.\ King
1998). Hence it seems not unlikely that the actual transfer rate is
close to the evolutionary mean.   

The theoretical value of the critical rate $\dmc$ depends on 
a number of not very well determined quantities. 
Most notably these are the hydrogen ionization temperature in the
disc, the relative disc thickness, the accretion 
efficiency $\eta$ and the albedo. A major justification for the
adopted normalisation comes from the fact that it matches the observed
location of the critical 
rate separating transient from persistent sources (van~Paradijs 1996).
The remaining uncertainty in the normalisation of
$\dmc$ does not affect the main results of this study, i.e. the existence
and location of the transient instability strip in the Hertzsprung
gap. This is a result of the steep gradient $\dd \dot M/\dd P$ of the transfer
rate in this phase.

The transient X--ray source GRO~J1655--40 is the first confirmed member of
the class of intermediate--mass BHXBs considered in this paper.  
GRO~J1655--40 is also close to the SXT instability strip and, given the
uncertainties in both the theoretical and observed values, might
actually be a system right in the instability strip, consistent with
its transient nature (e.g.\ Harmon et al.\ 1995, Tavani et al.\ 1996,
Levine et al.\ 1996). Implications of this interpretation are
discussed elsewhere (Kolb et al.\ 1997). Particular problems arise
from the apparent lack of persistently bright systems with similar
parameters in the Hertzsprung gap. It was suggested that these are not
seen in X--rays as their effective photosphere might radiate at much
longer wavelengths.  
Alternatively, GRO~J1655--40 could be one of the predicted systems in
the Hertzsprung gap with a stable accretion disc. It is by no means
clear if such systems do appear as persistently bright
sources. Instabilities in the accretion flow might cause dramatic 
changes in the effective photosphere, causing  variability in a given
waveband. Kolb et al.\ (1997) suggested such a mechanism also for
GRS~1915+105, the other X--ray source with apparent superluminal
motion.   

A second possible intermediate--mass BHXB is 4U~1543-47. Recently,
Orosz et al.\ (1998) determined its orbital period as 1.123~d and
found that the components are likely to be a 
$\simeq 7\msun$ BH and a $2.1-2.5\msun$ main--sequence donor. This 
places the system below the turn--off main--sequence period ($P_{\rm
TMS}$) line in Fig.~7 (dash--dotted; note that $P_{\rm TMS}$ as shown
in the equivalent 
Fig.~2 of Kolb et al.\ 1997 is slightly too small due to a calibration
error). Therefore 4U~1543-47 is, unlike GRO~J1655--40, in a phase of
Case A mass transfer where the transfer rate is determined by the
donor's nuclear timescale. With $M_2=2.3\msun$ and replacing $K$ in
(\ref{eq3}) by the nuclear expansion $\dd \ln R/\dd t$ on the main
sequence we estimate $\dot M_2 \simeq 4 \times 10^{-9}
\msun$~yr$^{-1}$, well below the critical rate KKS find for transient
behaviour in BHXBs (but slightly above the critical rate claimed for
neutron star systems, cf.\ KKB).  
A more detailed account of Case~A mass transfer BHXBs is in
preparation.

\bigskip

This work was partially supported by the U.K.\ Particle Physics and
Astronomy Research Council (PPARC). I am grateful to Hans Ritter and
Juhan Frank for useful comments, and to Andrew King for very helpful 
discussions and for improving the language of the manuscript. Comments
of the anonymous referee helped to improve the manuscript.


{}


\begin{thebibliography}{}

\bibitem{}
Cannizzo J.K., Chen W., \& Livio M. 1995, ApJ, 454, 880

\bibitem{}
Cannizzo J.K., Wheeler J.C., \& Ghosh P. 1982, in {\it Pulsations in
Classical and Cataclysmic Variable Stars}, ed.\ J.P.~Cox, C.J.~Hansen
(Boulder: Univ.\ of Colorado Press), p.~13

\bibitem{} 
Chen W., Shrader C.R., \& Livio M. 1997, ApJ, 491, 312

\bibitem{} 
De~Greve J.P. 1993, A\&AS 97, 527

\bibitem{}
Harmon B.A., Wilson C.A., Zhang S.N., Paciesas W.S., Fishman G.J.,
Hjellming R.M., Rupen M.P., Scott D.M., Briggs M.S., \& Rubin
B.C. 1995, Nature, 374, 703

\bibitem{}
Hjellming M.S. 1989, PhD thesis, University of Illinois

\bibitem{}
Hjellming R.M. \& Rupen M.P. 1995, Nature, 375, 464

\bibitem{}
Kalogera V., \& Webbink R.F. 1996, ApJ, 458, 301


\bibitem{}
King A.R. 1998, in 13th North American Workshop on Cataclysmic
Variables, ed.\ S.~Howell, E.~Kuulkers, C.~Woodward, (San~Francisco:
ASP), in press 

\bibitem{}
King A.R. 1988, QJRAS, 29, 1

\bibitem{}
King A.R., \& Ritter H. 1998, MNRAS, 293, L42

\bibitem{}
King A.R., Kolb U., \& Burderi, L. 1996, ApJ 464, L127 (KKB)

\bibitem{}
King A.R., Kolb U., \& Szuszkiewicz E. 1997, ApJ, submitted (KKS)

\bibitem{}
King A.R., Frank J., Kolb U., \& Ritter H. 1997, ApJ, 484, 844

\bibitem{}
King A.R., Frank J., Kolb U., \& Ritter, H. 1996, ApJ, 467, 761

\bibitem{}
Kippenhahn R., \& Weigert A. 1967, Zeitschrift f\"ur
Astrophysik, 65, 251

\bibitem{}
Kolb U. 1996, in {\em Cataclysmic Variables and Related Objects}, ed.\
A.~Evans, J.H.~Wood, IAU Coll.~158, (Dordrecht: Kluwer), p.~433

\bibitem{}
Kolb U., \& Ritter H. 1990, A\&A 236, 385

\bibitem{}
Kolb U., King A.R., Ritter H., \& Frank J. 1997, ApJ, 485, L33

\bibitem{}
Levine A.M., Bradt H., Cui W., Jernigan J.G.,
Morgan E.H., Remillard R., Shirey R.E., \& Smith D.A. 1996, ApJ, 469,
L33 

\bibitem{}
Lin D.N.C., Taam R.E. 1984, in {\it High Energy Transients in
Astrophysics}, ed.\ S.E.~Woosley, AIP Conf.\ Proc.\ 115, p.~83 

\bibitem{}
Mazzitelli I. 1989, ApJ 340, 249

\bibitem{}
Orosz J.A. \& Bailyn C.D. 1997, ApJ, 477, 876 

\bibitem{}
Orosz J.A., Jain R.K., Bailyn C.D., McClintock J.E., \& Remillard R.A.
1998, ApJ, in press


\bibitem{}
Pylyser E.H.P., \& Savonije G.J. 1988, A\&A 191, 57

\bibitem{}
Pylyser E.H.P., \& Savonije G.J. 1989, A\&A 208, 52

\bibitem{}
Ritter H. 1996, in Evolutionary Processes in Binary Stars, ed.\
R.A.M.J.~Wijers, M.B.~Davies M.B., C.A.~Tout, 
NATO ASI, Series C, Vol.~477, (Dordrecht: Kluwer), p.~223 

\bibitem{}
Ritter H., Zhang Z., \& Kolb U. 1995, in Cataclysmic Variables,
ed. A.~Bianchini, M.~Della~Valle, M.~Orio, (Dordrecht: Kluwer Academic
Publishers), Astrophysics and Space Science Library, 205, 479


\bibitem{}
Romani R.W. 1994, in {\em Interacting Binary Stars}, ed.\
A.~Shafter, ASP Conf.\ Series Vol.~56, 196


\bibitem{}
Salaris M., Dom\'inguez I., Garc\'ia-Berro E., Hernanz M., Isern J., \&
Mochkovitch R. 1997, ApJ, 486, 413

\bibitem{}
Schaller G, Schaerer D., Meynet G., \& Maeder A.S. 1992, A\&AS, 96, 269

\bibitem{}
Tanaka Y., \& Shibazaki N. 1996, ARAA 34, 607

\bibitem{}
Tavani M., Fruchter A., Zhang S.N., Harmon B.A., Hjellming R.N., Rupen
M.P., Bailyn C., Livio M. 1996, ApJ, 473, L103

\bibitem{}
Van Paradijs J. 1996, ApJ, 464, L139

\bibitem{}
Warner B. 1987, MNRAS, 227, 23 

\bibitem{}
Webbink R.F., Rappaport S., \& Savonije G.J. 1983, ApJ, 270, 678

\bibitem{}
White N.E., \& van~Paradijs J. 1996, ApJ, 473, L25

\end{thebibliography}
\end{document}